\documentclass[twocolumn]{aa}
\usepackage{graphicx}
\begin{document}
   \title{Evidence of TeV gamma-ray emission from the nearby 
starburst galaxy NGC 253}
	\titlerunning{TeV gamma-ray from NGC 253}


   \author{C.~Itoh\inst{1}\and 
R.~Enomoto\inst{2}\thanks{\email{enomoto@icrr.u-tokyo.ac.jp}}\and
S.~Yanagita \inst{1}\and T.~Yoshida\inst{1}\and 
T.~Tanimori \inst{3}\and
K.~Okumura \inst{2}\and
A.~Asahara \inst{3}\and
G.V.~Bicknell\inst{4}\and
R.W.~Clay\inst{5}\and
P.G.~Edwards\inst{6}\and
S.~Gunji\inst{7}\and
S.~Hara\inst{3,8}\and
T.~Hara\inst{9}\and
T.~Hattori\inst{10}\and
Shin.~Hayashi\inst{11}\and
Sei.~Hayashi\inst{11}\and
S.~Kabuki\inst{2}\and
F.~Kajino\inst{11}\and
H.~Katagiri\inst{2}\and
A.~Kawachi\inst{2}\and
T.~Kifune\inst{12}\and
H.~Kubo \inst{3}\and
J.~Kushida\inst{3,8}\and
Y.~Matsubara\inst{13}\and
Y.~Mizumoto\inst{14}\and
M.~Mori\inst{2}\and
H.~Moro\inst{10}\and
H.~Muraishi\inst{15}\and
Y.~Muraki\inst{13}\and
T.~Naito\inst{9}\and
T.~Nakase\inst{10}\and
D.~Nishida\inst{3}\and
K.~Nishijima\inst{10}\and
M.~Ohishi\inst{2}\and
J.R.~Patterson\inst{5}\and
R.J.~Protheroe\inst{5}\and
K.~Sakurazawa\inst{8}\and
D.L.~Swaby\inst{5}\and
F.~Tokanai\inst{7}\and
K.~Tsuchiya\inst{2}\and
H.~Tsunoo\inst{2}\and
T.~Uchida\inst{2}\and
A.~Watanabe\inst{7}\and
S.~Watanabe\inst{3}\and
T.~Yoshikoshi\inst{16}}
	\authorrunning{Itoh, Enomoto, Yanagita, Yoshida et al.}
   \offprints{R. Enomoto}

   \institute{
Faculty of Science, Ibaraki University,
Mito, Ibaraki 310-8512, Japan 
\and
Institute for Cosmic Ray Research, Univ. of Tokyo, Kashiwa,
Chiba 277-8582, Japan 
\and
Department of Physics, Kyoto University, Sakyo-ku, Kyoto 606-8502, Japan
\and
MSSSO, Australian National University, ACT 2611, Australia
\and
Department of Physics and Math. Physics, University of Adelaide, SA 5005, 
Australia
\and
Institute for Space and Aeronautical Science, Sagamihara, 
Kanagawa 229-8510, Japan
\and
Department of Physics, Yamagata University, Yamagata, Yamagata 990-8560, Japan
\and
Department of Physics, Tokyo Institute of Technology, Meguro-ku, Tokyo 152-8551, Japan
\and
Faculty of Management Information, Yamanashi Gakuin University, 
Kofu,Yamanashi 400-8575, Japan
\and
Department of Physics, Tokai University, Hiratsuka, Kanagawa 259-1292, Japan
\and
Department of Physics, Konan University, 
Hyogo 658-8501, Japan
\and
Faculty of Engineering, Shinshu University, Nagano, Nagano 380-8553, Japan
\and
STE Laboratory, Nagoya University, Nagoya, Aichi 464-8601, Japan
\and
National Astronomical Observatory of Japan, Mitaka, Tokyo 181-8588, Japan
\and
Department of Radiological Sciences, 
Ibaraki Prefectural University of Health Sciences,
Ibaraki 300-0394, Japan
\and
Department of Physics, Osaka City University, Osaka, Osaka 558-8585, Japan}

   \date{Published in A\&A 402, 443-455 (2003)}

\abstract{
TeV gamma-rays were recently detected from
the nearby normal spiral galaxy NGC~253 (Itoh et al., 2002).
Observations to detect the Cherenkov light images 
initiated by gamma-rays from the direction of NGC~253
were carried out in 2000 and 2001 over a total period of $\sim$150 hours.
The orientation of images in gamma-ray--like events is not consistent
with emission from a point source, and the emission region corresponds
to a size greater than 10~kpc in radius.
Here, detailed descriptions of the analysis procedures and techniques 
are given.

}

\maketitle

%

\section{Introduction}

NGC~253 is a very nearby ($d=2.5$\,Mpc) 
(de Vaucouleurs, \cite{Vaucouleurs}), normal spiral,
starburst galaxy.  Starburst galaxies are generally expected to have
cosmic-ray energy densities about hundred times larger than that of our
Galaxy (Voelk et al., \cite{Voelk}) due to the high 
rates of massive star formation
and supernova explosions in their nuclear regions. The star-formation
rates can be estimated from the far-infrared (FIR) luminosities,
and the supernova rates can be also inferred based on the assumption of an
initial mass function. Since the supernova rate of NGC~253 is
estimated to be about 0.05 - 0.2 yr$^{-1}$ 
(Mattila and Meikle, \cite{Mattila}, Antonucci and Ulvestad, 
\cite{Antonucci}, van Buren and Greenhouse, \cite{Buren}), a high cosmic-ray
production rate is expected in this galaxy.

We recently reported on the detection of TeV gamma-rays from
NGC~253 (Itoh et al. 2002).  Previous to this, the only evidence for
higher energy particles in a galaxy other than our own is for the
Large Magellanic Cloud (Sreekumar et al., \cite{sreekumar92}).  
Voelk et al. (\cite{VAB}) estimated
the gamma-ray fluxes (via neutral pion decay) from the nucleus of nearby
starburst galaxies. These values, however, were under the sensitivity
of the EGRET detector on the \textit{Compton Gamma-Ray Observatory}
(CGRO) and, indeed, EGRET observations resulted in very stringent
upper limits for the GeV emission from 
NGC~253 (Blom et al., \cite{Blom}, 
Sreekumar et al., \cite{sreekumar94}).

On the other hand, NGC~253 has an extended synchrotron-emitting halo
of relativistic electrons (Carilli et al., \cite{Carilli}). 
The halo extends to a large-scale height, 
where inverse Compton scattering (ICS) may be a more
important process for gamma-ray production than pion decay and
bremsstrahlung. The seed photons for ICS are expected to be mainly FIR
photons up to a few kpc from the nucleus, and cosmic microwave
background radiation at larger distances.

The OSSE instrument onboard the CGRO detected sub-MeV gamma-rays from
NGC~253 (Bhattacharya et al., \cite{Bhatt}). 
This emission is consistent with a model for ICS
of the FIR photons around the nucleus of the galaxy by
synchrotron-emitting electrons (Goldshmidt and Rephaeli, \cite{gold}), 
although it is difficult
to study the spatial distribution of the emission due to the limited
angular resolution of OSSE.

We observed NGC~253 with the CANGAROO-II telescope in 2000 and
2001, and detected TeV gamma-ray emission at high statistical
significance (Itoh et al. 2002).  This detection of TeV gamma-rays
from a normal spiral galaxy like our own has profound implications for
the origin and distribution of TeV cosmic-rays in our Galaxy.  In this
paper we describe in detail the observations and analysis of the TeV
gamma-rays from NGC~253.

\section{Observations}

\subsection{The CANGAROO-II telescope}

The CANGAROO (Collaboration of Australia and Nippon (Japan) for a
GAmma Ray Observatory in the Outback) air Cherenkov telescope is
located near Woomera, South Australia (136$^{\circ}$46'E,
31$^{\circ}$06'S, 220~m a.s.l.).  The telescope consists of a 10\,m
reflector and a 552 pixel camera.  It detects images of cascade showers
resulting from sub-TeV gamma-rays (and background cosmic rays)
interacting with the Earth's upper atmosphere.

The CANGAROO-II project is exploring the southern sky at gamma-ray
energies of 0.3$\sim$100~TeV.  Its predecessor, CANGAROO-I, used a
3.8\,m telescope (Hara et al., \cite{hara}), 
and detected TeV gamma-ray emission from
such objects as pulsar nebulae (PSR 1706-44 (Kifune et al., \cite{kifune95}),
the
Crab~(Tanimori et al., \cite{tanimori98a})), supernova remnants (SNR)
(SN1006~(Tanimori et al., \cite{tanimori98b}), 
and RX\,J1713.7$-$3946~(Muraishi et al., \cite{muraishi00})).
The 10\,m telescope of CANGAROO-II has been in operation since April,
2000, and has detected SNR RX\,J1713.7$-$3946 (Enomoto et al, 
\cite{enomoto_n})
and the active galactic nuclei Mrk 421 (Okumura et al., \cite{okumura02}).  
The
telescope has a parabolic optical reflector consisting of 114
composite spherical mirrors (80\,cm in diameter), made of carbon fiber
reinforced plastic (CFRP) (Kawachi et al., \cite{kawachi01}).  
The principal parameters
of the telescope are listed in Table~\ref{tab:mainpara}.
\begin{table}[htbp]
     \begin{tabular}{ll}
        \hline
        \hline
        Parameters & Values\\
        \hline
        \hline
	Location  & 136 $^\circ$E ,31$^\circ$S \\
	Height above sea level & 220~m \\
	Total diameter & 10~m \\
	Focal length   & 8~m \\
	Number of segmented mirrors  & 114\\
	Mirror diameter & 80~cm \\
	Mirror segment shape & Spherical\\
	Mirror alignment & Parabolic\\
	Mirror curvature & 16.4~m\\
	Mirror material & Plastic (CFRP)\\
        \hline
     \end{tabular}
  \caption[]{Principal parameters of the CANGAROO-II telescope.}
  \label{tab:mainpara}
\end{table}

The camera contains 552 pixels, each of which subtends an area of
0.115$^\circ \times$0.115$^\circ$.  Each pixel is a 1/2$"$
photomultiplier tube (PMT) (Hamamatsu Photonics R4124UV) with an air
light guide. The output signal is amplified by a high-speed IC (Lecroy
TRA402S) and split three ways for the ADC (analogue to digital
converter), TDC (time to digital converter), and the scalers.
The scaler is a special circuit which records
the number of hits greater than the threshold ($>$ 2.5 photoelectrons) 
of individual PMTs within 700\,$\mu$sec
(Kubo et al.,\cite{kubo}). The scalers were triggered by a clock (1 Hz),
and these data were recorded every second.

\subsection{Pointing direction and observation}

The telescope was pointed at the center of NGC~253, the J2000
coordinates of which are (RA, Dec)=($11.888^{\circ},-25.288^{\circ}$).
NGC~253 was observed from October~3 to November~18, 2000 and from
September~20 to November~15, 2001, with the CANGAROO-II telescope.
The observations were carried out on clear nights during moon-less
periods.  Periods of 1.5 hours after sunset and 1.5 hours before sunrise
were avoided.  Each night was divided into two or three periods, i.e.,
ON--OFF, OFF--ON--OFF, or OFF--ON observations.
ON-source observations were timed to contain the meridian passage of
NGC~253, which culminates at a zenith angle of $\sim 6^{\circ}$.  The
observation times are summarized in Table \ref{tab:obs_period}.
\begin{table}[htbp]
     \begin{tabular}{lll}
        \hline
        \hline
        Observation Date & T$_{on}$(min) & T$_{off}$(min)\\
        \hline
        03-Oct. -- 18-Nov. 2000 & 2297 & 2245\\
        20-Sep. -- 15-Nov. 2001 & 2567 & 2401\\
        \hline
        Total & 4864 & 4646\\
        \hline
        \hline
     \end{tabular}
  \caption[]{Summary of the observation periods.}
  \label{tab:obs_period}
\end{table}
In total, $\sim$4800 minutes of ON-source observations and a similar amount of 
OFF-source observations were carried out.

\subsection{Hardware Trigger}

\begin{figure}[htbp]
  \centering
    \includegraphics[width=6cm]{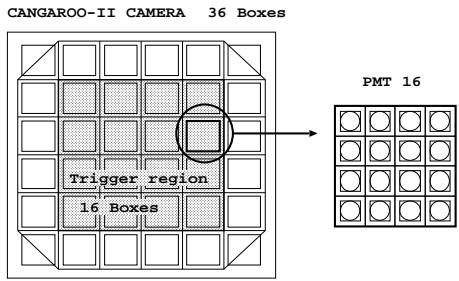}
    \caption{Pixel arrangement of the CANGAROO-II camera.
        The thick Box~(1.84$^\circ \times$1.84$^\circ$)
        is the trigger region. The camera consists of 36 boxes.
        Each box contains 16 PMTs, each 1/2$"$ in diameter.
        In total, 552~PMTs are installed.
        Each pixel subtends 0.115$^\circ \times$0.115$^\circ$,
        defined by the light guide.}
    \label{fig:camera}
\end{figure}

The pixel arrangement of the CANGAROO-II camera is shown in 
Fig.~\ref{fig:camera}. 
The trigger-region is defined by the inner 
1.84$^\circ \times$1.84$^\circ$ square, which
contains 256 PMTs in 16 boxes.
The event trigger requires:
\begin{enumerate}
\item More than three pixels to be hit inside the trigger region.
The threshold for each pixel was set at approximately 2.5
photoelectrons (p.e.).
\item More than one box with a charge-sum exceeding $\sim$10~(p.e.).
\end{enumerate}

\section{Data analysis}

\subsection{Calibration}

The data were calibrated using a LED (Light Emitting Diode) 
light source located at the
center of the 10\,m mirror, $\sim$8~m from the camera (Kabuki et al., 
\cite{kabuki}).
A quantum-well type blue LED (NSPB510S, $\lambda \sim 470$\,nm,
Nichia Corporation, Japan)
was used, and illuminated with an input pulse of $\sim$ 20 nsec width.
A light diffuser was placed in front of the LED in order to obtain a
uniform yield on the focal plane.  The main purpose of this
calibration was field flattening.  The relative gain of each pixel was
adjusted according to the mean pulse height of all pixels.  The second
purpose was to adjust the timing of each pixel with respect to the
mean timing for all pixels.  Time-walk corrections (adjusting the
earlier triggering of larger pulses that arises from a fixed trigger
threshold) were also carried out, based on the data.  This calibration
was done run by run.

\subsection{ADC conversion factor}

In order to compare the simulated and observed spectra, the energy
scale must be calibrated, i.e.\ a conversion factor from the ADC value
to the absolute energy is required.  First, we checked the cosmic ray
event rate. Under the assumption that $\sim 100$~ADC counts
corresponded to a single photoelectron, the cosmic-ray rate roughly
agreed.  Using a Monte-Carlo simulation (described later) of
cosmic-ray protons, we further studied this ADC conversion factor
(Hara, \cite{hara2}).  We analyzed the relation between the total ADC counts
and the total number of pixel hits.  From this correlation we
determined this factor to be 92 $^{+13}_{-7}$ [ADC ch/p.e.].  This
agreed with the results of a study of the Night Sky Background (NSB)
rate.

\subsection{Pixel Selection}

\begin{figure}[htbp]
  \centering
    \includegraphics[width=8cm]{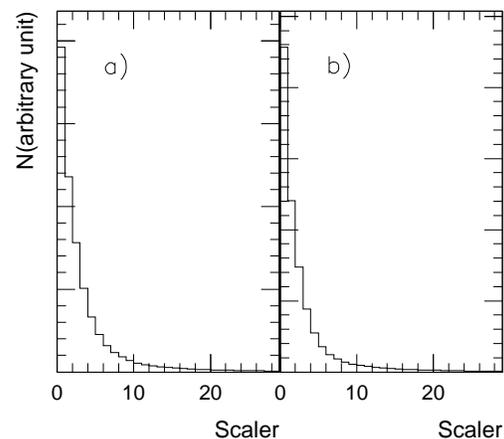}
    \caption{Scaler distribution: a) 2000-data
        and b) 2001-data, where "N" denotes
the number of hits per scaler count (horizontal axis). 
The rate at which each pixel
        exceeds the $\sim$ 2.5 p.e.\ threshold was monitored
        for 700~$\mu $s each second.}
    \label{fig:sclcut}
\end{figure}

Occasionally, individual pixels display anomalously high count rates.
The trigger rate of each pixel is monitored by a scaler every second.
This information enabled us to remove these ``hot'' pixels from
further analysis.  The scaler distributions obtained from 2000- and
2001-data are shown in Figs.~\ref{fig:sclcut} a) and b), respectively.
In 2000, due to the influence of artificial lighting from the
detention centre several km away, the hit rate was significantly higher
than that of 2001, and so a slightly higher cut value was adopted for
the 2000 data.  There were no stars brighter than a magnitude of 5.6
in the field of view (FOV) of the camera during these observations.
However, the effects of fainter stars passing through the FOV of a
pixel were expected to be removed by this PMT rate cut.  This was
confirmed using data from other observations which had brighter stars
in the FOV of the camera.

\subsection{Clustering}

\begin{figure}[htbp]
  \centering
    \includegraphics[width=10cm]{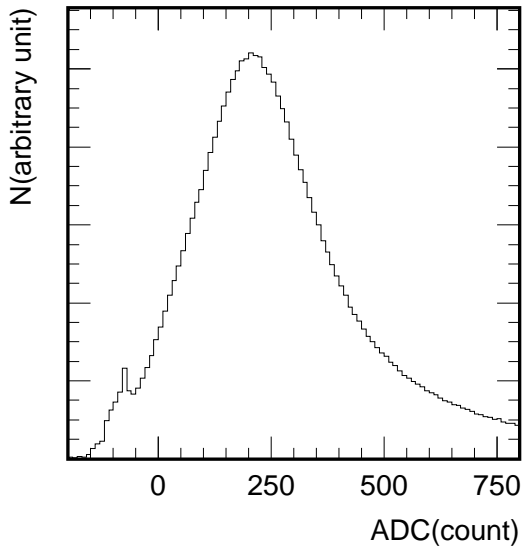}
    \caption{ADC distribution for pixels, after pedestal levels had been
subtracted, where "N" denotes the number of hits per 10-ADC count. 
The small peak in negative region
is due to the electronics undershoot.
}
    \label{fig:adc}
\end{figure}

The purposes of the pre-selection were to remove noisy pixels affected
by the NSB and any period affected by cloudy conditions from the
observation data.  Here, we used ``t$n$a'' logic (threshold
$n$-adjacent, where $n$ is the number of adjacent PMTs required to
have triggered).  The threshold was fixed at around 300 ADC count
(approximately 3.3~p.e.).  The distribution of ADC is shown was in
Fig.~\ref{fig:adc}.
The hardware threshold was located at $\sim$200 ADC counts. 
Therefore, the cut at 300~ADC counts ($\sim$3.3 p.e.) is reasonable. 
This is the simplest and most powerful method to reject pixels affected by NSB.

\begin{figure}[htbp]
  \centering
    \includegraphics[width=10cm]{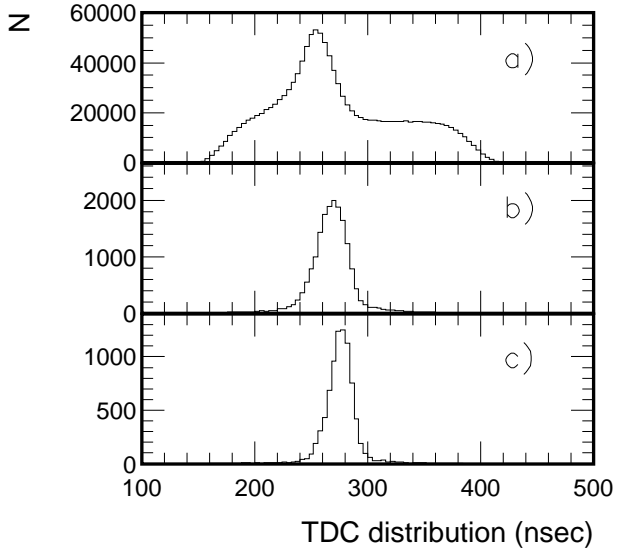}
    \caption{TDC distributions: a) for all pixels, b) for pixels 
which satisfy t3a-clustering and do not satisfy t4a, and 
c) for pixels which satisfy t4a-clustering and do not satisfy t5a.
"N" denotes the number of hits per TDC count, which were normalized to 1~nsec.
The horizontal axis is the TDC count.}
    \label{fig:tdist}
\end{figure}

After this selection, those clusters with more than $n$ adjacent hits
were selected.  As $n$ increased, the TDC distribution became
cleaner, as shown in Fig.~\ref{fig:tdist}.  The mean event timing was
located at around 300~TDC counts (1~count~=~1~nsec).  Those events
uniformly distributed between 200 and 400~nsec are considered to be
due to NSB photons.  From Fig.~\ref{fig:tdist}, we selected a cut of
$n=4$.  We also cut pixels with $|$TDC-300$| > 40$~nsec.

\begin{figure}[htbp]
  \centering
    \includegraphics[width=10cm]{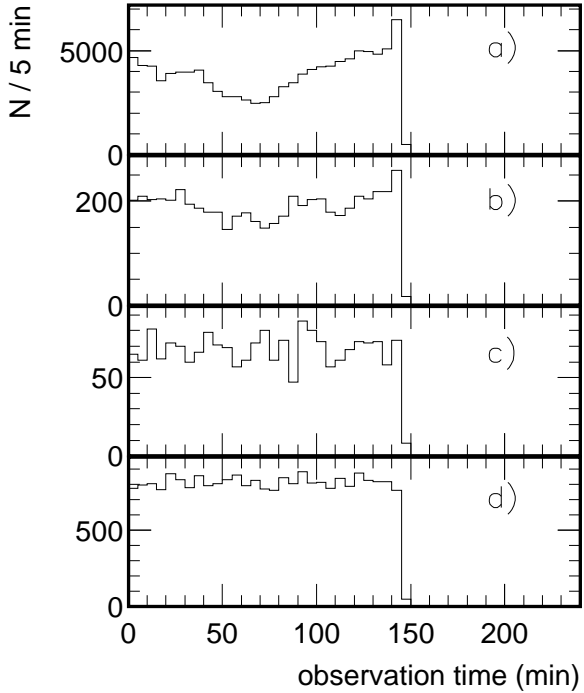}
    \caption{Shower rates: a) hardware trigger rate;
b) shower rate for events satisfying t3a-clustering, but not
satisfying t4a;  
c) shower rate for events satisfying t4a-clustering, but not satisfying
t5a; and
d) shower rate for t4a-clustering. 
The vertical axis "N" denotes the number of events per 5 min.
The horizontal axis is the time in minutes from the start of 
the observation. 
}
    \label{fig:srate}
\end{figure}

After this cut, the event rate which satisfies t4a-cluster and does
not satisfy t5a is shown in Fig.~\ref{fig:srate} c).
Although the raw trigger rates were not stable, due to changes in the
background light level as the telescope pointing changed, the shower
rate became stable when t4a-selection was applied, as can be seen in
Fig.~\ref{fig:srate} c).  The shower rate after the t4a-clustering is
shown in Fig.~\ref{fig:srate} d).
From then on, the events with t4a-clustering were selected.

\begin{figure}[ht]
  \centering
    \includegraphics[width=10cm]{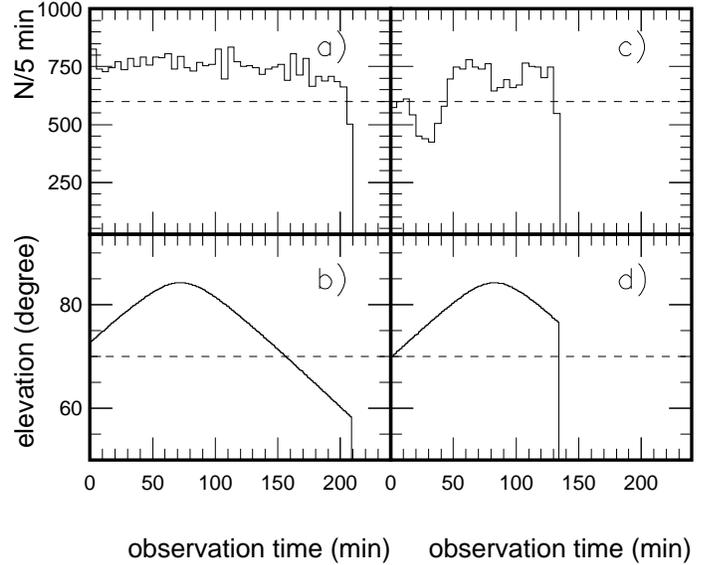}
    \caption{ Shower rate under a) good conditions and c) poor conditions.
 The horizontal axis is the time (minutes) from the start of an
 observation. The vertical axis ("$N$") is the number of events per 5 minutes.
 The histogram is the shower rate, and the dashed line is the cut position, 
 i.e., 600 events per 5 minutes were required for the data to be accepted.
 Plots b) and d) are the elevation angle distributions versus time.
 Data was accepted above an elevation angle of 70$^\circ$.}
    \label{fig:cloud_cut}
\end{figure}

\begin{figure}[ht]
  \centering
    \includegraphics[width=10cm]{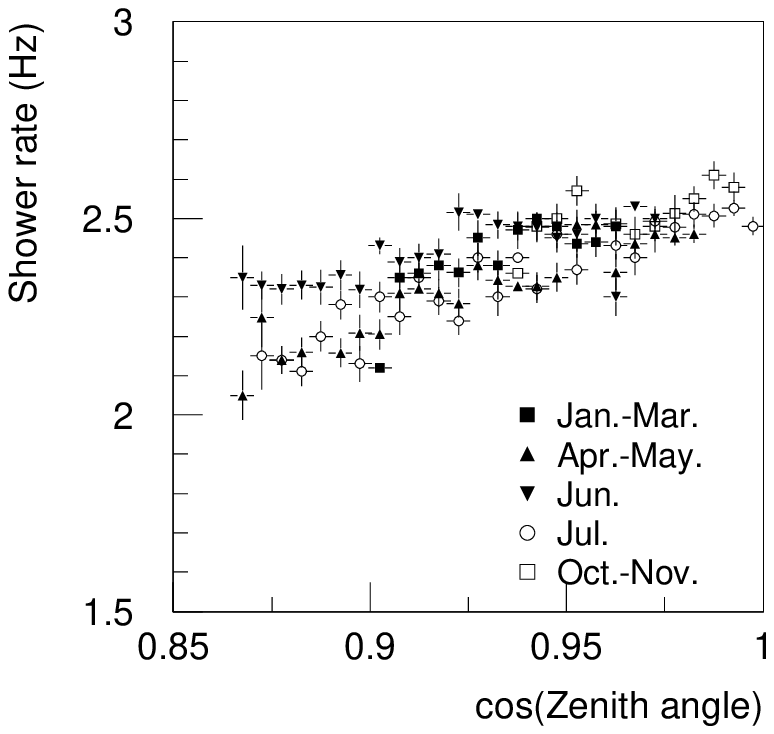}
    \caption{Shower rates versus the cosine of the zenith angle.  }
    \label{fig:srate_avx}
\end{figure}

Using these shower rate plots, we were also able to remove any cloudy period
during an observation. 
Examples of these plots for good and bad conditions are shown in 
Figs.~\ref{fig:cloud_cut} a) and c), respectively. 
The cloudy periods detected in this manner perfectly matched the 
observing conditions described in the experimental log.
The cut line (the dashed line in Fig.~\ref{fig:cloud_cut}) was set at 2.0~Hz,
corresponding to 600 events per five minutes.
The shower rate of the data passing these cuts was very stable over 
all observations, in all seasons, except for the expected zenith 
angle dependence. 
The stability of the shower rate in 2001 is shown in Fig.~\ref{fig:srate_avx}.
This stability implies that the degradation of the mirror reflectivity
was small.
Over a short period in 2000, the camera system was exposed to strong
background light from the township of Woomera, and the shower rate was 
observed to change by $\sim$20\%. This corresponds to a change in the
energy scale of 10\% for the $E^{-2.7}$ cosmic-ray spectrum.
A systematic error of 10\% is included in 
the energy determination as a result.
Also, we rejected any events with zenith angles greater than 20 degrees,
as shown in Figs.~\ref{fig:cloud_cut} b) and d). 

These observations were carried out in the southern hemisphere during spring.
Near sunrise, the humidity increased, and dew started to form
on the surfaces of the mirrors.  
These effects could also be detected by this shower-rate study,
and we eliminated these periods from any further analysis.

After these pre-selections, the data remaining for analysis were
accumulated, as summarized 
in Table~\ref{tab:selected_data}.
\begin{table}[htbp]
     \begin{tabular}{llll}	
        \hline	
        \hline	
        Observation period & T$_{on}$(min) 
	& T$_{off}$(min)& T$_{on}$/T$_{off}$\\
        \hline
        3 Oct. - 18 Nov. & 1301 & 969 &1.34\\
        2000&&&\\
        20 Sep. - 15 Nov. & 1658 & 1448 &1.15\\
        2001&&&\\
        \hline
        Total & 2959 & 2417 & 1.22\\
        \hline
        \hline
     \end{tabular}
  \caption[]{
	Summary of data remaining after pre-selection cuts.}
  \label{tab:selected_data}	
\end{table}

\subsection{Image Analysis}

\subsubsection{Monte-Carlo Setups}

Simulations of electromagnetic and hadronic showers in the atmosphere
were carried out using a Monte-Carlo simulation code based on
GEANT3.21 (\cite{GEANT}).  In this code, the atmosphere was divided into
80 layers of equal thickness ($\sim 12.9 {\rm g/cm}^{2}$)
(Enomoto et al., \cite{enomoto_ast}).  
Each layer corresponds to less than a half
radiation length.  The dependence of the results on the number of
layers was checked by increasing the number of layers, and was
confirmed to be less than a 10$\%$ effect.  The lower energy threshold
for particle transport was set at 20~MeV, which is less than the
Cherenkov threshold of electrons at ground level.
Most Cherenkov photons are emitted higher in the atmosphere, at lower
pressure and a higher Cherenkov threshold.  The geomagnetic field at
the Woomera site was included in the simulations (a vertical component
of 0.520\,G and a horizontal component of 0.253\,G directed
6.8$^{\circ}$ east of south).

In order to save CPU time, Cherenkov photons were tracked in the
simulations only when they were initially directed to the mirror
area.  The average measured reflectivity of 80$\%$ at 400\,nm and its
wavelength dependence (Kawachi et al., \cite{kawachi01}) 
and the measured PMT quantum
efficiency were multiplied using the Frank-Tamm equation to derive the
total amount of light and its wavelength dependence.  A
Rayleigh-scattering length of $2970(\lambda /400 {\rm nm})({\rm g/cm}^{2})$
(Baum and Dunkelman, \cite{baum1955}) 
was used in transport to the ground.  No Mie
scattering was included in this study.  The contribution of Mie
scattering is thought to be greatest at the 10--20$\%$ level; we
therefore consider this study to have uncertainties of at least this
level.  When Rayleigh scattering occurred, we treated it as absorption.

\begin{figure}[htbp]
  \centering
    \includegraphics[width=10cm]{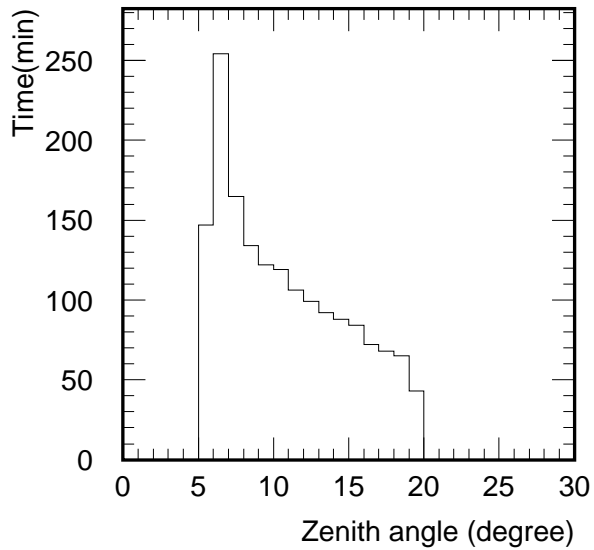}
    \caption{Zenith-angle distribution after event selection.
	The vertical axis  is the observation time in minutes.
        The mean is 11.1$^{\circ}$.}
    \label{fig:zenith}
\end{figure}

Finally, the simulated electronic noise was added and the timing responses were
smeared using a Gaussian of 4~ns (1$\sigma$). We also added NSB
photons, conservatively selecting to double Jelley's value of
$2.55\times 10^{-4} {\rm erg/cm^{2}/s/sr}$ (430-550 nm) 
(Jelley, \cite{jelley}). 
Electronics saturation was also taken into account.  The zenith angle
distribution was obtained from an ON-source run, and is shown in
Fig.~\ref{fig:zenith}.
The above distribution was input to the event generator of the
Monte-Carlo simulation. 
We generated gamma-rays between 100~GeV and 10~TeV
assuming a (Crab-like) power-law spectral index of $-$2.5.

\begin{figure}[htbp]
  \centering
    \includegraphics[width=8cm]{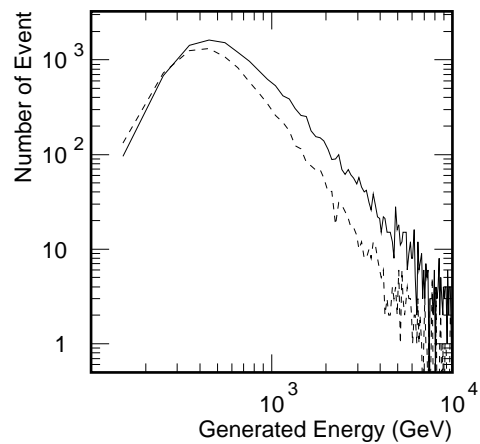}
    \caption{Distribution of energies for accepted events from
	the  Monte-Carlo gamma-ray simulation; the solid line
        was obtained for a E$^{-2.5}$ spectrum and the dashed line 
        for a E$^{-3.0}$ spectrum.}
    \label{fig:e_gen}
\end{figure}

The energy spectrum for simulated 
events passing the pre-selection cuts are shown in Fig.~\ref{fig:e_gen}.
From this figure, we obtained the threshold energy for the gamma-ray
detection to be $\sim$ 500~GeV for a $E^{-2.5}$ spectrum
and $\sim$ 400~GeV for a $E^{-3.0}$ spectrum,
after pre-selection cuts.
These simulated gamma-ray events were used together with
observed events from OFF-source runs to
determine the cut values in order to optimize the gamma-ray signal. 

\subsubsection{Data analysis}

\begin{figure}[htbp]
  \centering
    \includegraphics[width=8cm]{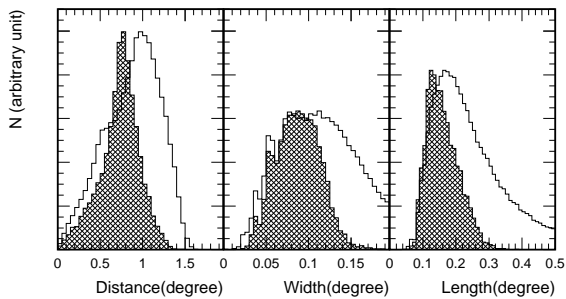}
    \caption{Image parameter distributions.
	The blank histograms were obtained from OFF-source observations.
	The hatched areas are the distributions for gamma-rays
	from our Monte-Carlo simulations.}
    \label{fig:image}
\end{figure}

We first calculated the standard image parameters: \textit{Distance}, 
\textit{Width}, and \textit{Length} (Hillas, \cite{hillas}).
The distributions of these parameters are shown in Fig.~\ref{fig:image}.
In order to check the background shapes of image parameters, we carried out
a simulation of cosmic-ray proton events, generating  protons (only)
between 500~GeV and 10~TeV from a differential $E^{-2.7}$ spectrum.
The mean elevation angle of OFF-source observations was assumed.
The distributions of the resulting image parameters were checked and found
to be roughly consistent with those obtained by the OFF-source runs.

\begin{figure}[htbp]
  \centering
    \includegraphics[width=10cm]{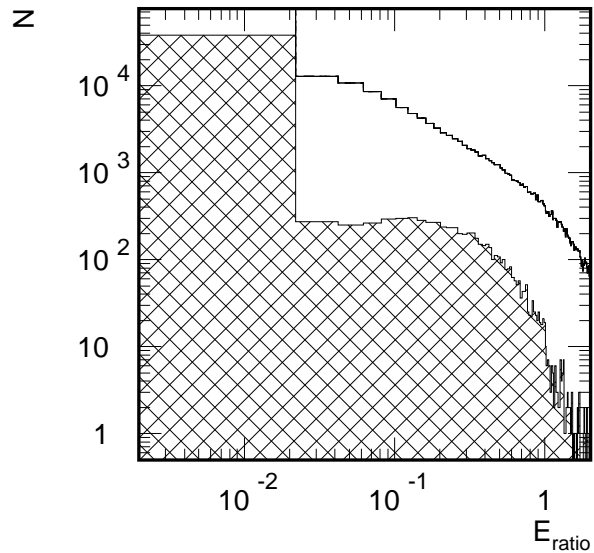}
    \caption{$E_{ratio}$: Ratio between the ADC counts of the maximum 
energy cluster and that of the remaining clusters. 
"N" denotes the number of events.
The blank histogram
was obtained from an OFF-source run and the hatched histogram from
the gamma-ray simulations.}
    \label{fig:eratio}
\end{figure}

We cut events with a \textit{Distance} of less than 0.5$^{\circ}$ or
greater than 1.2$^{\circ}$.  The \textit{Width} and \textit{Length}
were used as a Likelihood ratio, which is described later.  We defined
$E_{ratio}$ as the sum of the ADC counts outside the main cluster in
the image, divided by the sum of the ADC sum inside the main (maximum
energy) cluster.  Gamma-ray events are predicted by simulations to be
typically a single cluster, and thus have low values of $E_{ratio}$.
The distribution of $E_{ratio}$ is shown in Fig.~\ref{fig:eratio}.
We rejected events with $E_{ratio} > 0.1$.
This cut helped to reduce the cosmic-ray backgrounds. 
Also, less-energetic events, with a total ADC count of less than 3000
($\sim$33 p.e.), were rejected in order to improve the 
$\alpha$ (image orientation angle) resolution.

The acceptance of gamma-ray--like events was evaluated
using the Likelihood-ratio (Enomoto et al., \cite{enomoto_ast} 
and \cite{enomoto_pro}). 
Probability Density Functions (PDFs) were derived for
both gamma-ray and cosmic ray initiated events.
The PDFs for gamma-rays were obtained from simulations,
while those for cosmic rays were obtained from OFF-source data. 
Histograms were made of  \textit{Length} and \textit{Width} 
using both data sets; these distributions were then normalized to unity. 
The probability (L) for each assumption was thus obtained by multiplying 
PDF(\textit{Width}) by PDF(\textit{Length}). 
In order to obtain a single parameter, and also to normalize it to unity, 
we used the Likelihood-ratio:
\begin{equation}
Likelihood-ratio=\frac{L(gamma-ray)}{L(gamma-ray)+L(proton)}.
\end{equation}

\begin{figure}[htbp]
  \centering
    \includegraphics[width=10cm]{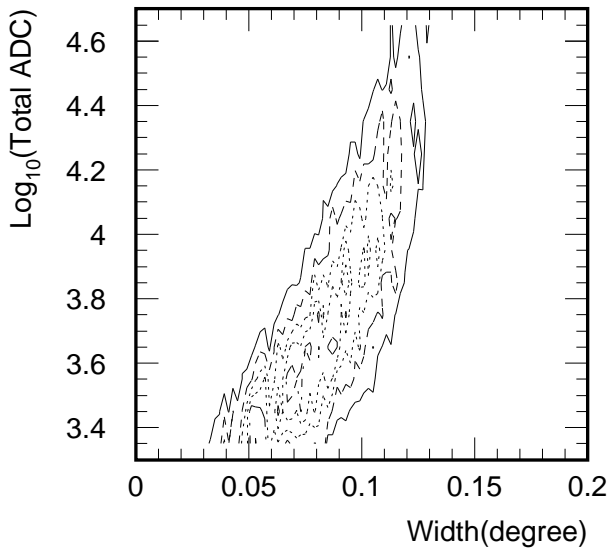}
    \caption{Correlation between the \textit{Width} and the logarithm of 
total ADC counts obtained from the gamma-ray simulations.}
    \label{fig:ene_wid}
\end{figure}

We took care to take account of the energy dependences of 
these two image parameters.
An example of the \textit{Width} parameter is shown in Fig.~\ref{fig:ene_wid}.
In order to correct for the energy dependence, we made 2D-histograms, i.e., 
the shape parameter versus the logarithm of the total ADC counts, 
and calculated PDFs for an energy independent acceptance of signal events.
Here, we did not use either the \textit{Distance} or \textit{Asymmetry} 
parameter, as these parameters are source-point dependent.
If the gamma-ray emitting region is broader than 
that of a point source, these parameters will deviate from
that of point source.
In the Monte-Carlo simulations for the PDF determination, 
we used the point-source assumption.

\begin{figure}[htbp]
  \centering
    \includegraphics[width=10cm]{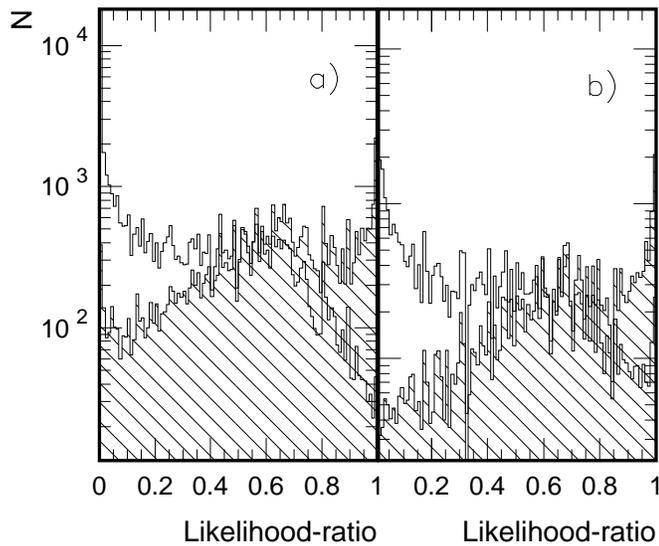}
    \caption{Likelihood Distribution: a) for 2000 data
	and b) for 2001 data.
"N" denotes the number of events.
	The blank histograms are OFF-source data and the 
	hatched areas are from the gamma-ray simulations.}
    \label{fig:like_dist}
\end{figure}

We selected gamma-ray--like events using the Likelihood-ratio. 
The Likelihood-ratio distributions are shown in Fig.~\ref{fig:like_dist}. 
The signal peaks at a Likelihood-ratio of 1 and the background at 0.

\begin{figure}[htbp]
  \centering
    \includegraphics[width=8cm]{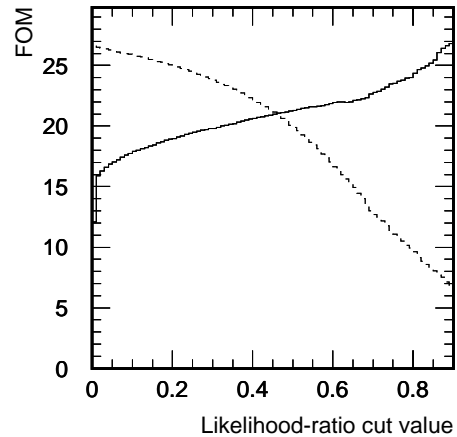}
    \caption{FOM (figure of merit) vs Likelihood-ratio cut for the 
combined-data 
is shown by the solid line. Also shown is the acceptance vs 
Likelihood-ratio cut (the dashed line).}
    \label{fig:best}
\end{figure}

We then investigated the figure of merit (FOM) using these data
in order to maximize the statistical significance of the 
gamma-ray signals from NGC~253.
At various cut locations, the signal of the Monte-Carlo simulation and 
the OFF background entry were obtained. The FOM was defined as the former 
value divided by the square root of the latter value. 
The FOM versus Likelihood-ratio cut values are plotted in Fig.~\ref{fig:best}.
The figure suggests that higher cut values lead to a higher
statistical significance, albeit with a loss in the gamma-ray acceptance.
As a compromise between the acceptance and the FOM, 
we opted to adopt a value of 0.4 for the Likelihood-ratio cut, noting
that there was only a small change in FOM between 0.2 and 0.6.

\subsection{Further Hot Pixel Rejection}
\label{sec:hot}

Pixels occasionally have anomalously high trigger rates, often 
due to enhanced starlight or man-made light in the FOV of the pixel,
or to small discharges between the light-guide and 
the photo cathode (Kabuki et al., \cite{kabuki}), or 
to electrical noise in the associated circuitry.
Although these are generally random, small pulse-height signals, the 
high pixel trigger rates can have the affect
of increasing the camera trigger rate.
When randomly triggered during a real event trigger, these ``hot pixels''
rarely form a cluster. 
Their effect was thought to be reduced after the pre-selection
clustering cuts.
However, it is possible that outlying hot pixels surviving 
the pre-selection cuts deform the shapes of the shower images.
Such effects could significantly smear the $\alpha$ distribution for
gamma-ray events.
In fact, the $\alpha$ distributions of the OFF-source runs were
observed to be deformed from the Monte-Carlo prediction.

In order to flatten these distributions, we removed hot pixels.
First we looked at the hot pixel map for events passing selection cuts.
This enabled the hottest pixels to be identified and removed.
We then looked at the scaler counts for the remaining pixels.
In the same way, some of the hottest pixels were removed.
Finally, we tested pixels iteratively to find out whether they deformed
the $\alpha$ distributions of OFF-source run events for clusters having
a center of gravity around the pixel being investigated.
For the 2000~data, 12.3$\%$ of the pixels were removed by
these operations. 
The same procedure was carried out for the 2001~data, 
and  9.7$\%$ of pixels were removed.

After removing those hot pixels, we checked the Monte-Carlo simulation
precisely, and verified that these procedures did not result in
any deformation of the image parameters, including $\alpha$.

Most of hot pixels were located around the edges of the camera.
This occurred as PMTs with high trigger rates, which were
identified early during camera testing, and were deliberately moved
from inside the trigger region to the outer edge of the camera
to minimize the effect on the hardware trigger.

In order to check whether small pulse-height random-noise signals
were removed by these operations, we loosened the clustering to t3a,
which should be more sensitive to these backgrounds.
Similar plots were obtained, which confirmed that they were still
consistent with the Monte-Carlo predictions.
With t4a clustering and these procedures, we concluded that the random
noises were removed successfully by this operation.
	
\subsection{Results}

\begin{figure}[htbp]
  \centering
    \includegraphics[width=8cm]{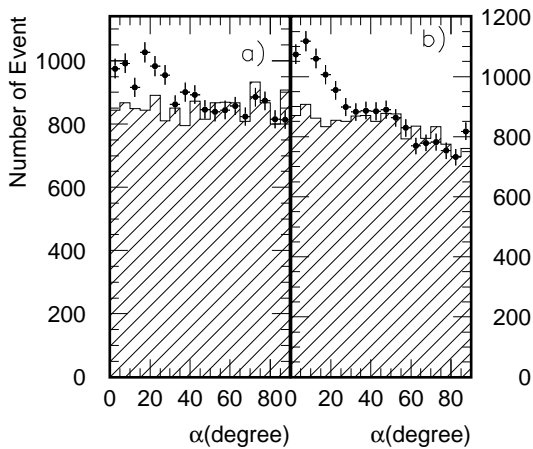}
    \caption{The distributions of $\alpha$ obtained for 
        a) 2000 data and b) 2001 data.
	The points with error bars were obtained from ON-source data.
	The shaded histogram is from the OFF-source data.}
    \label{fig:alpha1y}
\end{figure}

The resulting $\alpha$~distributions are shown in Fig.~\ref{fig:alpha1y}.
The excesses at $\alpha
<$ 30$^\circ$ are 742.5$\pm$104.6 events (7.1$\sigma$) for 2000 and
933.1$\pm$106.2 events (8.8$\sigma$) for 2001, respectively.  By
combining the two years of data, the total excess was 
found to be 1651.9$\pm$149.2
(11.1$\sigma$).  Here, we used an Likelihood-ratio cut of 0.4.  
The $\alpha$ cut 
at 30$^{\circ}$ was larger than expected for a point source.
This cut value was used in the previous detection of gamma-rays 
from RX~J1713.7$-$3946, which was found to have an extended nature 
(Enomoto et al, \cite{enomoto_n}). 
The distortion of the $\alpha$ spectrum in both ON- and OFF-source
runs was observed in 2001 (Fig. \ref{fig:alpha1y}-b).
The ON and OFF spectra, however, agreed well for $\alpha > 30^o$,
even including the normalization factor, which is described in
section \ref{sec:crab}.
These appeared in higher $\alpha$ regions, and were considered to be
due to hot channels which remained even after the rejection procedure
described in the previous section. 
Most of these hot channels were located outside of the trigger region.
In order to keep the high efficiency
of the analysis, the hot channels which had less affect on the deformation
of the $\alpha$ spectrum were not removed.
For a point source, simulations predict that an $\alpha$ 
cut of between 15 and 20$^{\circ}$ should optimize the signal.
The simulations indicate that 73\% of the excess from a point source should 
have $\alpha < 15^{\circ}$. 
We tested $\alpha$ cuts from 15 to 35$^{\circ}$ in 5$^{\circ}$ steps 
in order to maximize the excess for this source, and
found that 30$^{\circ}$ was best. The statistical significance 
of the signal thus needs these 5 trials to be taken into account.
The final significance remained greater than 10$\sigma$. 
The signal rates were (743$\pm$105)~events/ 1301min = 0.57$\pm$0.08 for
2000 and (933$\pm$106)~event/ 1658min = 0.56$\pm$0.06 for 2001,
consistent within the statistical errors.
The average event rate overall was 0.56$\pm$0.05 /min.

\subsection{Various Checks}

\begin{figure}[htbp]
  \centering
\includegraphics[width=6cm]{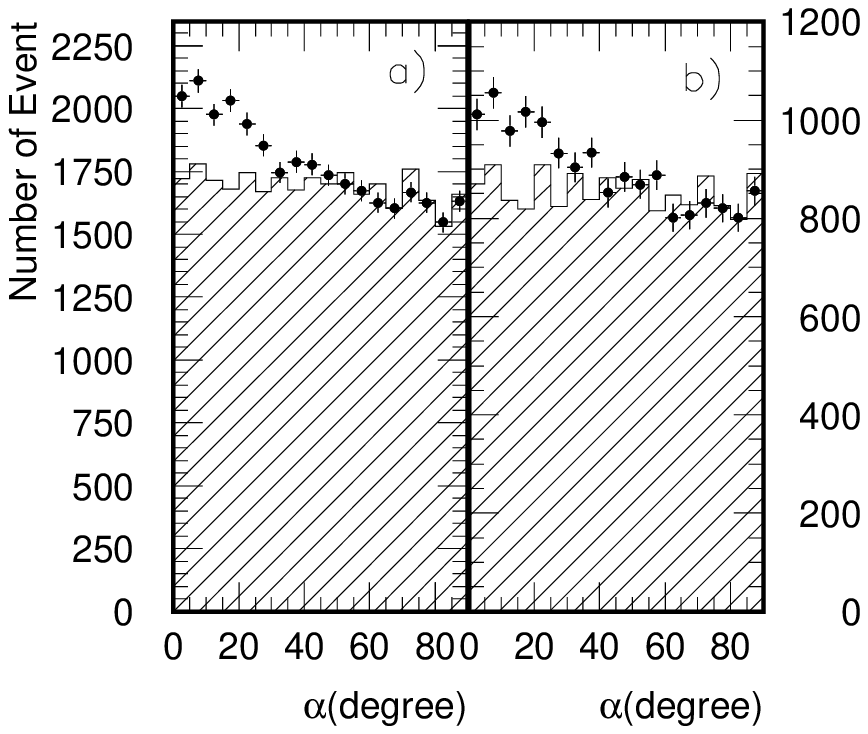}\includegraphics
[width=6cm]{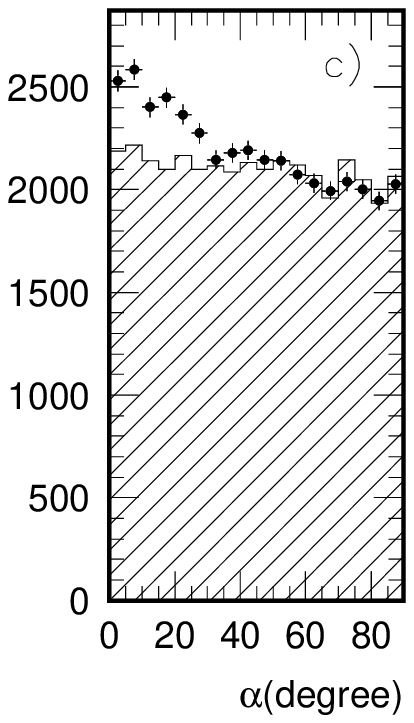}
    \caption{$\alpha$ distributions obtained with a) the loose cut 
	($L > 0.4$), b) a tighter cut ($L > 0.6$), and c)
        ``standard'' cuts.
	The points with error bars were obtained from ON-source data.
	The shaded histogram represents OFF-source data.}
    \label{fig:alpha2y}
\end{figure}

We investigated the effect of raising the Likelihood-ratio cut, 0.6, 
i.e., applying a tighter cut. 
The results for the combined (2000 and 2001) data set
are shown in Fig.~\ref{fig:alpha2y} b).
The excess for the tight cut is
823.8$\pm$105.6 (7.8$\sigma$),
somewhat lower, as expected due the reduced acceptance.

In order to verify the Likelihood method, we checked the results of
the ``standard'' analysis, using an acceptance cuts of 
0.05$<$ \textit{Width} $<$0.15 and 0.05$<$ \textit{Length}
$<$0.3.  The $\alpha$ distributions are shown in
Fig.~\ref{fig:alpha2y} c).  The excess is 1696.7$\pm$165.9
(10.2$\sigma$), with signals of 811.7$\pm$119.0~events (6.8$\sigma$) and
916.7$\pm$115.4~events (8.0$\sigma$) in 2000 and 2001, respectively.  
As expected, the standard analysis confirms the statistical significance
of the detection, though at a lower level than the more powerful
Likelihood-ratio method.

\subsection{Effective Area}

\begin{figure}[htbp]
  \centering
    \includegraphics[width=8cm]{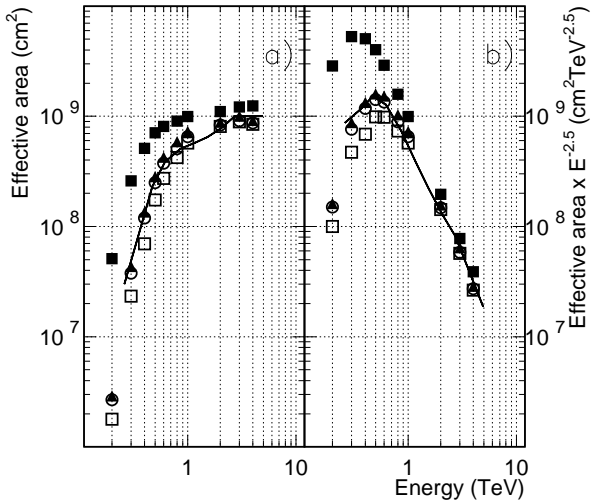}
\caption{Effective area versus energy; a) Effective areas after
the pre-selection (the black squares), those after the distance cut
(the black triangles), those after the Likelihood-ratio cut (the blank
circles), and those with $\alpha < 15^o$ (the blank squares).
The effective area for the Whipple telescope is shown by the line (after
the distance cut). b) The effective areas multiplied by $E^{-2.5}$ are
shown in order to indicate the threshold of the CANGAROO telescope.}
\label{fig:effarea}
\end{figure}

The effective areas for this analysis is shown in Fig. \ref{fig:effarea}.
We compared them with that of Whipple
(Fig. 6 in Mohanty et al.\ \cite{moh98}),
which are the effective areas after clustering and Distance cut,
i.e., before image parameter cut. 
Our effective area agreed with Whipple, even 
with the energy dependences.
According to Fegan (\cite{feg96}), the threshold
of Whipple telescope was the same as ours.

\section{Discussion}

\subsection{Crab Analysis}
\label{sec:crab}

\begin{figure}[htbp]
  \centering
    \includegraphics[width=8cm]{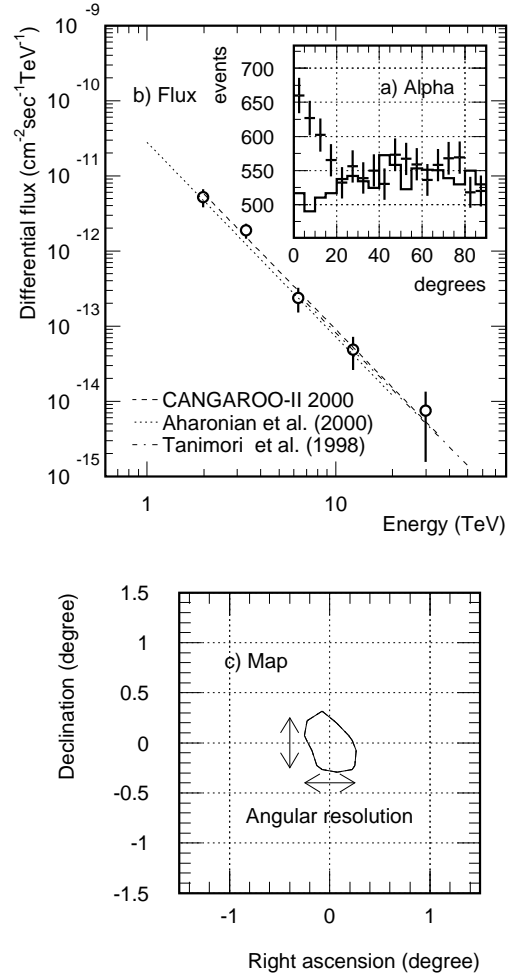}
    \caption{
Results of the Crab analysis.
The differential flux obtained for the Crab nebula is shown in b), 
	together with previous observations.
  	The points with error bars were obtained by this experiment.
        The dotted line is the HEGRA result 
        and the dashed line is the CANGAROO-I result.
        The insert (a) is the $\alpha$ distribution. The points
        with error bars are obtained from ON-source data and the histogram is
        from OFF-source data. The lower plot (c) is a significance
        map. The 65\%-contour is drawn with the our estimated angular
        resolution (the arrows: $\pm 1 \sigma$).}
    \label{fig:crab2000_flux}
\end{figure}

We observed the Crab nebula in November and December, 2000, in 
order to check our energy and flux
determination. The Crab nebula has a power-law
spectrum over a wide energy range (Aharonian et al., \cite{aharonian00}, 
Tanimori et al., \cite{tanimori98b}).
The elevation angle ranged from 34$^{\circ}$ to 37$^{\circ}$.
Approximately 10 hours of ON- and OFF-data were used for analysis.
The energy threshold was estimated from simulations to be $\sim$2~TeV.
The excessed number of events was 405 $\pm$ 59 (6.8$\sigma$), as shown in Fig.
\ref{fig:crab2000_flux} a).
The differential flux, shown in
Fig.~\ref{fig:crab2000_flux}-b), is consistent with 
results from HEGRA (Aharonian et al., \cite{aharonian00}) 
and CANGAROO-I~(Tanimori et al., 
\cite{tanimori98a})
observations.
We conclude that our energy and flux estimations are correct.  In
addition, we derived a cosmic-ray spectrum from background events and
compared it with the known cosmic-ray flux.  From these checks, the
systematic uncertainty for the absolute flux estimation was found to
be within 10$\%$.

The $\alpha$ plot for the Crab nebula, shown as an insert
(Fig.~\ref{fig:crab2000_flux} a), is consistent with the point-source
assumption.  
The OFF-$\alpha$ spectrum was again not flat.
The average image positions were different from the zenith-angle
observation, i.e., they were centerized because the shower max position
was higher in altitude. The reflective index of air was smaller there,
which resulted in a smaller Cherenkov angle.
In addition, there is a bright star close to the Crab position.
In order to avoid a high trigger rate,
we displaced (0.25 degrees) the telescope's tracking center away from the
center position of the Crab nebula. 
The average hit region in the camera was different from the observation
of NGC 253.
Thus, a different $\alpha$ deformation
(due to the hot channels) occurred in this case.
The 65\% contour obtained from the significance map is
shown in the lower plot Fig.~\ref{fig:crab2000_flux}c). 
The arrows are the estimated angular
resolution ($\pm 1\sigma$ = $\pm 0.25 ^\circ$).  Note that this is
larger than that of the NGC~253 analysis (0.23$^{\circ}$), due to
the zenith-angle dependence.  The Crab was observed at zenith
angles of around 56$^\circ$, whereas NGC~253 was observed at around 6$^\circ$.  
The center of the significance map corresponds to the Crab pulsar, confirming
that our pointing and angular resolution are consistent with our Monte-Carlo
simulations.

\begin{figure}[htbp]
  \centering
    \includegraphics[width=10cm]{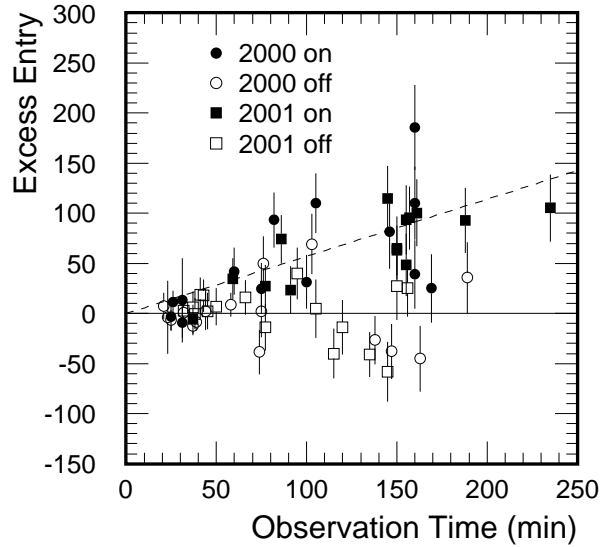}
    \caption{Excesses versus observation times
(after cloud cut) for all runs.
	The blank marks were obtained by an OFF-source run,
	and the filled marks by ON-source runs.
	The dashed line is the average flux.}	
    \label{fig:runbyrun}
\end{figure}

Miscellaneous checks were carried out, as described here.  The
agreements between the ON and OFF $\alpha$ distributions in a region
greater than 30 degrees were compared with the observation times
listed in Table. \ref{tab:selected_data}. The result is
$\frac{N_{on}/N_{off}}{T_{on}/T_{off}}=1.02$ .  
The signal rate for each individual observation was
calculated, and the results are plotted in Fig.~\ref{fig:runbyrun}.
They are consistent with the average flux described above.

\subsection{Systematic uncertainty in the  Energy determination}

There were uncertainties in determining the energy scale from the total
ADC counts.  The ADC conversion factor was 92$^{+13}_{-7}$, as
previously described.  The mirror reflectivity also had some
uncertainty, in both its value (averaged over the whole mirror) and 
its time dependence.  A measure of the latter could be made 
from month-by-month shower rates.  Also,  Mie
scattering was not taken into account in our Monte-Carlo
simulations.  Considering all of these effects, we estimated the systematic
uncertainty in the energy determination to be within 15$\%$ (bin to
bin) and 20$\%$ (overall).  These are also consistent with the Crab
analysis results described previously.

\subsection{Differential Flux}

\begin{figure}[htbp]
  \centering
    \includegraphics[width=10cm]{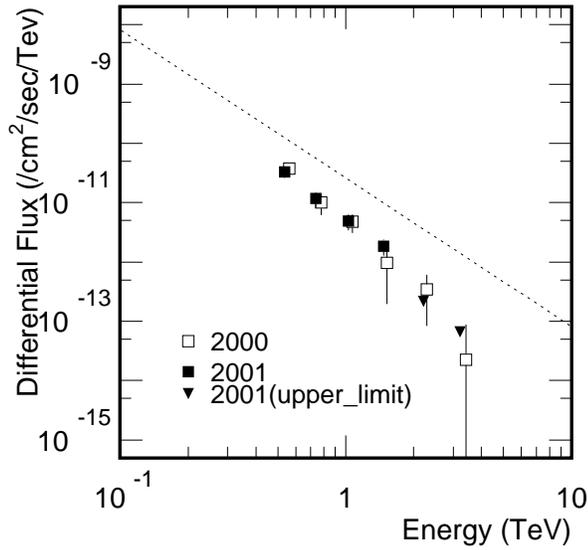}
    \caption{Differential fluxes for the 2000 and 2001 data.
        The blank squares represent data from 2000, and the
	filled squares are from the 2001 data.
	The triangles are the upper limits.
	The dotted line is that of Crab nebula for a reference.}
    \label{fig:flux_upp}
\end{figure}

The energy spectra of TeV gamma-ray sources generally have
a power-law nature.
Therefore, we used the ${\rm log_{10}} (Energy)$ scale in binning events
to determine the spectrum for NGC~253, rather
than energy, itself.
The energy for each event was assigned as a function
of the total ADC counts, where the relation between the energy and the
total ADC counts was obtained from simulations.  The excess (gamma-ray) 
events were observed between 0.5 and 3~TeV.  We
divided this ${\rm log_{10}}(Energy)$ range using equipartition.  The number
of binnings is 6. In Fig.~\ref{fig:flux_upp}, the derived spectra
for both 2000 and 2001 are plotted, and are seen to be consistent
with each other.

\begin{figure}[htbp]
  \centering
    \includegraphics[width=10cm]{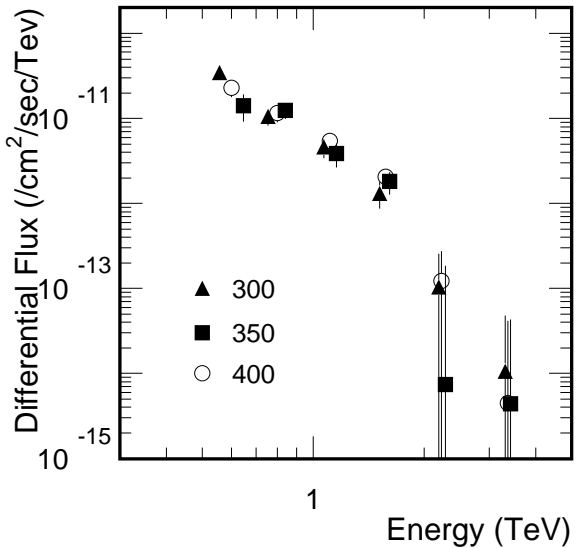}
    \caption{ADC threshold dependence of the differential fluxes.
       }
    \label{fig:plot_th}
\end{figure}

The systematic uncertainties were estimated as follows.  
The background light sources, such
as stars and artificial light may have a significant effect on the
estimation of the differential flux determinations.  These backgrounds
affect each pixels pulse-height distribution.  
Small contributions (Poisson distributed) would be added to the
signal in each pixel. 
In order to study the significance of this effect, 
we varied the ADC threshold from 300 (default), to 350 and 400 
(corresponding to 3.3, 3.8, and 4.3 p.e., respectively).  
The differential flux was obtained  
for each case and plotted in Fig.~\ref{fig:plot_th}.
The total excesses of signals varied 1652$\pm$149 events (300 count threshold),
1429$\pm$154 events (350), and 1034$\pm$160 events (400),
respectively.  
The acceptances were (as expected) strongly dependent on the threshold
value, but the differential fluxes were stable, as shown in
Fig.~\ref{fig:plot_th}.

\begin{figure}[htbp]
  \centering
    \includegraphics[width=10cm]{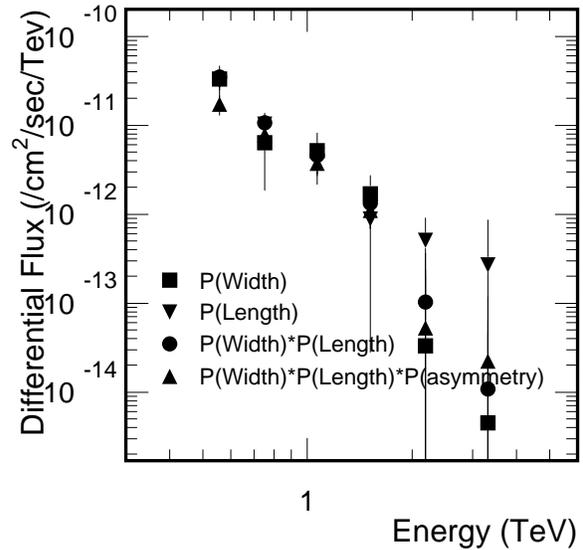}
    \caption{Differential fluxes based on 
various assumptions concerning the Likelihood-ratios.
       }
    \label{fig:plot_para}
\end{figure}

As a further check that the excess events are due to gamma-rays,
the following tests were also made.  We re-calculated the 
Likelihood-ratios 
by first adding $asymmetry$, then removing $length$, and finally,
removing $width$. 
This was done to check whether only one parameter had an unduly large 
effect on the final signal.  The resulting fluxes are shown in
Fig.~\ref{fig:plot_para}.
A deviation was observed when we added $asymmetry$ to the Likelihood-ratio.
$Asymmetry$ could only be calculated under the assumption of a
point source at the center of the FOV.  
As we had already concluded that the source was extended,
the fact that the use of $asymmetry$ reduces the significance
of the signal is not unexpected, and confirms that it is
not a useful parameter for diffuse radiation.
The other Likelihood-ratios were in agreement with each other,
supporting our conclusion that the excess is due to diffuse gamma-ray 
emission. A 
summary of flux changes in this parameter study is listed in 
Table~\ref{tab:para}.
\begin{table}[htbp]
     \begin{tabular}{lllllll}
        \hline
        \hline
        Energy (TeV) & 0.56 &0.75 &1.07 &1.52&2.19&3.32\\\hline
        Flux change (\%) & 23.4 & 12.0 & 8.8 & 17.4 & 72.8 & 88.5\\
        \hline        \hline
     \end{tabular}
  \caption[]{Summary of the flux changes in the parameter study at 
each energy binnings.}
  \label{tab:para}
\end{table}
The mean energies listed in this table were obtained by averaging
the generated energies of the accepted events in the
ADC binnings in the above-described Monte-Carlo simulation.

\begin{figure}[htbp]
  \centering
    \includegraphics[width=10cm]{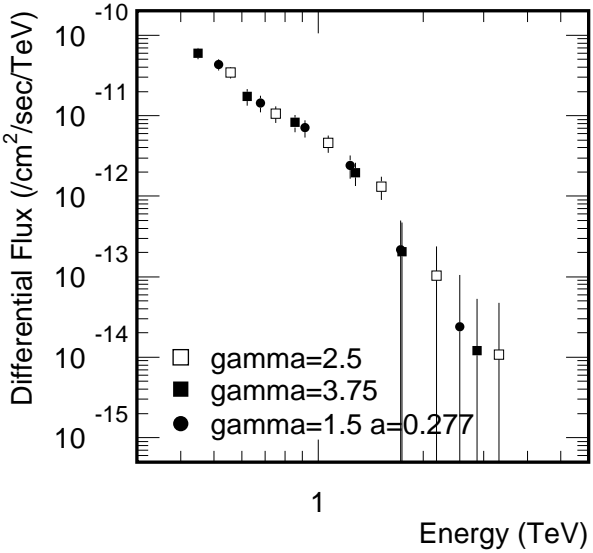}
    \caption{Differential flux estimated by the various energy
spectra inputs in the Monte-Carlo simulation. The blank squares were
obtained by $E^{-2.5}$, the black squares by $E^{-3.75}$, and the black
circles by the exponential cutoff function described in the text.
       }
    \label{fig:iteration}
\end{figure}

In order to derive the spectrum of NGC~253 self-consistently,
we adopted the following method.
The above spectrum was derived by using acceptances
derived from simulations in which a $E^{-2.5}$ spectrum
was assumed.  If we fit the fluxes with a differential power-law
spectrum, we obtained an index of $-$3.7$\pm$0.3.  We then iteratively
used this value in simulations to re-derive the spectrum.
This process rapidly converged at an index of $-$3.75$\pm$0.27.  
The differential fluxes estimated with $E^{-2.5}$ input
and that with $E^{-3.75}$ are shown in Fig.~\ref{fig:iteration}.
Both showed the same best fit spectrum.

An extrapolation of this power-law spectrum to lower energies deviates 
greatly from the measured fluxes and upper limits 
(Blom et al., \cite{Blom}, Sreekumar et al., \cite{sreekumar94}); a
turn-over below the TeV region clearly exists.  
Physically plausible functions exist with a turn-over include spectra
$\propto E^{-\gamma}e^{-E/E_{max}}$ and 
$\propto E^{-\gamma}e^{-\sqrt{E}/a}$.  
Although the former is typically used for the spectra of gamma-rays originating
from $\pi^o$ decay, the value of $\gamma$ should
be greater than 2.0 according to the present acceleration theories,
in contradiction with the measurements at lower energies.  The
latter form is typical for an Inverse Compton origin. 
We fitted a spectrum of this form with the EGRET upper limits.  
The best fit with this function 
gave $a=0.28$ with a reasonable $\chi^2$ value. We tried to generate
events with this spectral input to derive the differential flux
again. These are shown in Fig.~\ref{fig:iteration} (the black
circles). The flux determination is very stable over a range of assumptions
for the Monte-Carlo inputs of the energy spectrum. We also carried out an
iteration with this function, and confirmed the convergence to be
good. Finally, we selected this to be the Monte-Carlo energy spectrum.

The systematic errors were estimated by varying the Likelihood-ratio cut 
values, as described previously. They are listed in Table. \ref{tab:err}.
\begin{table}[htbp]
     \begin{tabular}{lllllll}
        \hline
        \hline
        Energy (TeV) & 0.52 &0.68 &0.92 &1.23&1.73&2.56\\\hline
        Sys. error (\%) & 29.8 & 23.4 & 34.8 & 23.9 & 32.6 & 67.9\\
        dE/E (\%) & 36.7 & 35.2 & 33.7 & 32.9 & 30.6 & 32.3\\
        \hline        \hline
     \end{tabular}
  \caption[]{Systematic errors and energy resolutions 
of each energy binning.}
  \label{tab:err}
\end{table}
These values are larger than those in Table \ref{tab:para}. We,
therefore, concluded to use these as the systematic errors.  Also
shown are the energy resolutions in each bin, which were obtained
from simulations on an event-by-event basis.  These errors are
dominated by the core distance uncertainties.  From here on, the flux
errors in the figures are the square root of the quadratic sum of the
statistical and systematic errors.

\begin{figure}[htbp]
  \centering
    \includegraphics[width=10cm]{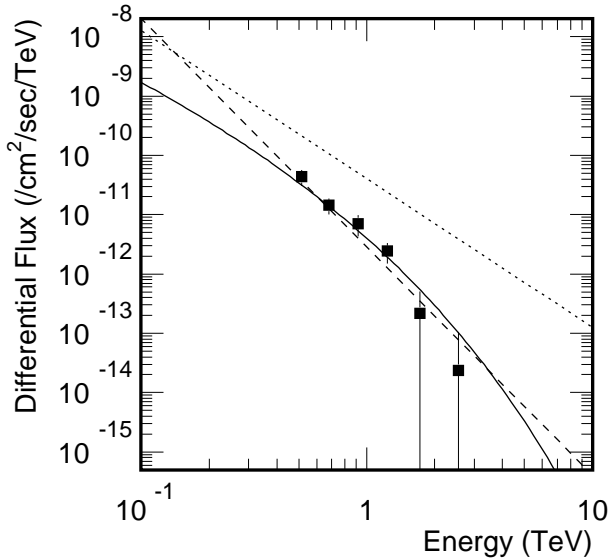}
    \caption{Combined differential fluxes.
	The dotted line is that of Crab nebula observations.
 	The other lines are the fitting results.
	The dashed line is that of a power-law.
	The solid curve is that with an exponential cutoff. }
    \label{fig:fit_repo}
\end{figure}

The combined flux is shown in Fig.~\ref{fig:fit_repo} and 
Table~\ref{tab:diff}.
Around 1~TeV, the flux is about one order lower than in the Crab nebula,
which is indicated by the dotted line in Fig.~\ref{fig:flux_upp}
(Tanimori et al., \cite{tanimori98a}).

\begin{table}[htbp]
     \begin{tabular}{llll}
        \hline
        \hline
 Log ADC region & $<$Energy$>$ & Flux \\
  & (TeV) & (/cm$^{2}$/sec/TeV) \\
        \hline
3.6--3.8 & 
	0.52 &  (3.51$\pm$0.72$\pm$1.04)$\times 10^{-11}$ \\
3.8--4.0 &
	0.68 &  (1.09$\pm$0.34$\pm$0.26)$\times 10^{-11}$ \\
4.0--4.2 &
	0.92 &  (6.40$\pm$1.69$\pm$2.23)$\times 10^{-12}$ \\
4.2--4.4 &
	1.23 &  (2.05$\pm$0.78$\pm$0.49)$\times 10^{-12}$ \\
4.4--4.6 &
	1.73 &  (2.65$\pm$2.83$\pm$0.87)$\times 10^{-13}$ \\
 4.6--4.8 &
	2.56 &  (2.31$\pm$8.07$\pm$1.57)$\times 10^{-14}$ \\
        \hline
        \hline
     \end{tabular}
  \caption[]{Energy binnings and differential fluxes.
The first errors are statistical and the second ones are systematic.}
  \label{tab:diff}
\end{table}

The goodness of fit for the various spectra were characterized
by the $\chi^{2}$ values. The results for various fittings are as follows:

\begin{eqnarray*}
\frac{dF}{dE}=(2.85\pm 0.71)\times 10^{-12}\times 
(E/{\rm 1 TeV})^{(-3.85\pm 0.46)} ~{\rm [cm^{-2}s^{-1}TeV^{-1}]},
\end{eqnarray*}
\begin{eqnarray}
\chi^{2}/DOF=2.1/4,
\end{eqnarray}
\begin{eqnarray*}
\frac{dF}{dE}=a e^{\sqrt{E_0}/b} 
(E/E_0)^{-1.5}e^{-\sqrt{E}/b}
~{\rm [cm^{-2}s^{-1}TeV^{-1}]},
\end{eqnarray*}
\begin{eqnarray*}
a=6 \times 10^{-5}~{\rm [cm^{-2}s^{-1}TeV^{-1}]},~E_0=0.0002~{\rm TeV},
\end{eqnarray*}
\begin{eqnarray}
b=0.25 \pm 0.01~[\sqrt{\rm TeV}],
\chi^{2}/DOF=1.8/5,
\label{eq3}
\end{eqnarray}
where the second formula is based on the Inverse Compton scattering
formula when we constrain the flux at 3/4 of the EGRET upper limit at
$E_0$=0.0002 TeV. 
Due to this, the two parameters in Eq. (\ref{eq3}), 
$a$ and $E_0$, were fixed
to those values.  
Here we assumed the power law index of the incident electron energy spectrum
to be 2.0.
A better $\chi^2$ was obtained for this fit 
compared to the single power-law fit.
In order to explain our flux and that of EGRET simultaneously, 
Eq. (\ref{eq3}) is one of reasonable choices.

\subsection{Morphology}

\begin{figure}[htbp]
  \centering
    \includegraphics[width=14cm]{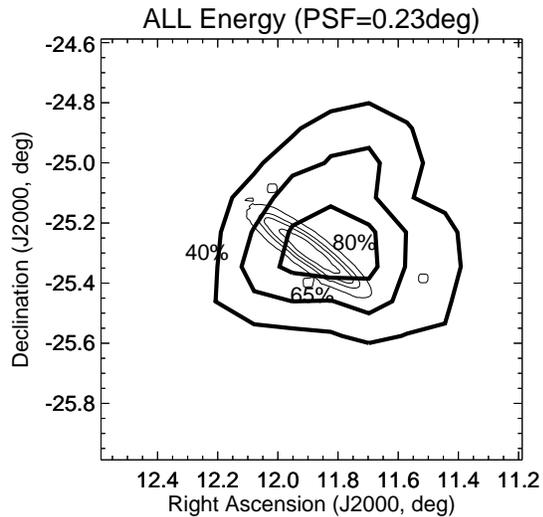}
    \caption{Significance map obtained by this experiment, shown 
	by the thick contours. The thin contours are optical image by DSS2.}
    \label{fig:significancemap}
\end{figure}

\begin{figure}[htbp]
  \centering
    \includegraphics[width=10cm]{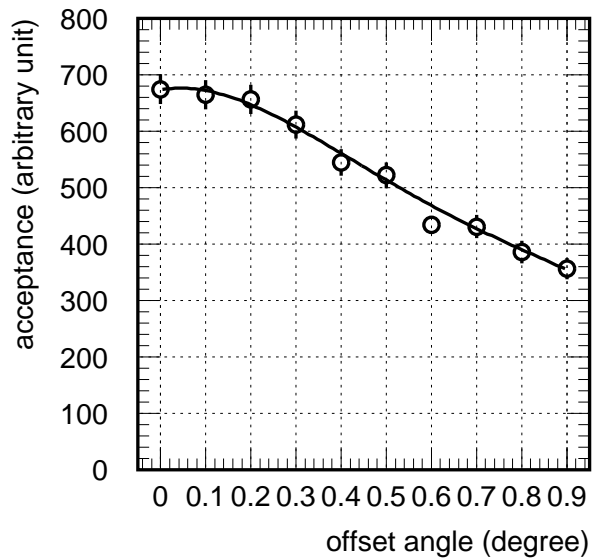}
    \caption{Efficiency versus offset angle of the gamma-ray
source position from the center of the field of view.}
    \label{fig:epsoff}
\end{figure}

The thick contours in Fig.~\ref{fig:significancemap} represent the
source morphology obtained from our observations. 
These contours were obtained from the so-called significance map and
are, therefore, not exactly speaking a morphology.  
The significance map was made from the
distribution of the detection significance determined at each point,
based on the assumption that each point in turn was a point-source position.
The significance was obtained from the difference in the $\alpha$ plot
(ON- minus OFF-source histogram) divided by the statistical errors.
The angular resolution of this method was estimated to be
$0.23^{\circ}$ (1$\sigma$ is a 68$\%$ confidence level).  The
telescope pointed at the center of NGC~253.  Also shown by the thin
contours is an optical image obtained by DSS2 (second version of
Digital Sky Survey).  The ``Significance'' is proportional to the intensity only
when the acceptance and the background level are uniform in the full
FOV.  The detection efficiency is dependent on the offset angle of
the assumed source position from the centre of the field of view,
as shown in Fig.~\ref{fig:epsoff}.

We checked the effect of the background light.  The optical
magnitude of 
NGC~253 has magnitudes of $m_B \sim 8$ and $m_{vis} \sim 7.1$. 
Even if it was concentrated on one point, the background level
due to this was lower than our sensitivity (Stars fainter than
a magnitude of 5 could not be detected in either the scaler
or ADC data).  We also note that the lower cut on
\textit{Distance} was 0.5$^{\circ}$, and so pixels around
the center of NGC~253 were generally not used for the analysis.  
When observing NGC~253, the
brightest stars in the FOV have magnitudes of 5.6.  
However, we observed some effects
from a group of faint stars (each of magnitude $\sim$6) in 
observations of another target, which deformed the shapes of
the parameter distributions.  These effects
are believed to be removed by the hot pixel rejection algorithm,
described in section~\ref{sec:hot}.  This was demonstrated in an
analysis of the Crab nebula data. 
Although the visual magnitude of the Crab nebula is 8.4, a
bright star (magnitude 3.1) is located within the FOV of our camera. 
Despite this, we were able to derive a significance map consistent 
with the other measurements (Fig.~\ref{fig:crab2000_flux}).  
Because we cannot rule out the possibility 
that ``hot pixels'' may deform the significance map, we can not 
definitely derive the morphology of the gamma-ray emitting regions
from observations with only a single telescope.

\begin{figure}[htbp]
  \centering
    \includegraphics[width=10cm]{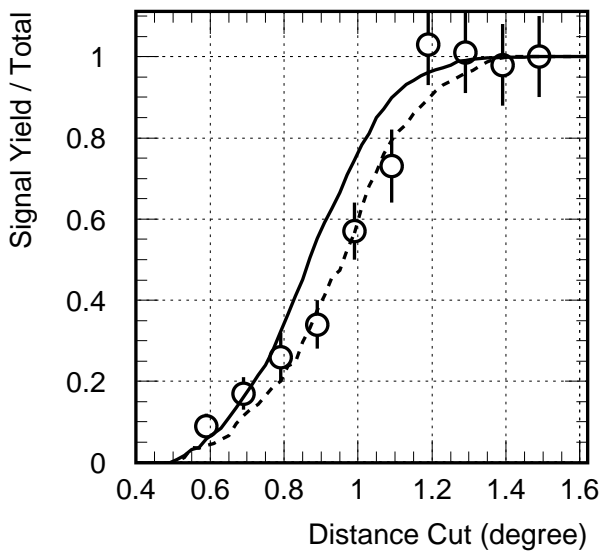}
    \caption{Signal yield as a function of the upper range of the 
        \textit{Distance} cut. The yield was normalized to the total excess.
	The points with error bars were obtained by this experiment.
	The solid line was obtained from a Monte-Carlo simulation 
	with a point source assumption.
	The dashed line is the case for a ``diffusion angle'' of
	0.4$^{\circ}$.}
    \label{fig:yield}
\end{figure}

Fig.~\ref{fig:yield} shows the acceptance of
gamma-ray--like events as a function of the \textit{Distance} upper cut values 
(minimum cut is $0.5^{\circ}$). 
From this figure, we tried to estimate the spatial extent of the
gamma-ray emission.  We proceeded by fixing the minimum cut
value of \textit{Distance} at $0.5^{\circ}$ (as used in
previous Whipple and CANGAROO analyses).  The maximum cut value
was varied between 0.6 and 1.5, and the excess events for each cut were
plotted by the points with error bars.
 
We checked the consistency between the experimental data and the
source diffusion assumptions.  At first, the solid line in
Fig.~\ref{fig:yield} was obtained by a Monte-Carlo simulation with a
point-source assumption.  The observed distribution is clearly broader 
than this.

From the significance map (Fig.~\ref{fig:significancemap}), the
correlations between the orientations of the TeV emission and optical image
were calculated and the standard deviation of the long
axis was 0.37$^{\circ}$ and of the short axis was 0.24$^{\circ}$, respectively.
The long axis
was inclined by $+$30$^{\circ}$ from the horizontal axis. This
is slightly larger than that of the optical image; however, we do not
believe that the difference is significant.  
An analysis of the map of the number
excess events yielded similar results.  We then carried out a
Monte-Carlo simulation based on various assumptions.  We varied the extent
of the emitting region by smearing the gamma-rays' incident angle with a
Gaussian.  Our data are consistent with "diffusion angles" of between
0.3-0.6$^{\circ}$.  The case of 0.4$^{\circ}$, shown in
Fig.~\ref{fig:yield} by the dashed line, is consistent with the
observation.  However, it is necessary to wait for 
future stereo observations (CANGAROO-III and H.E.S.S.) 
before any quantitative estimates of the extent of the emission region
can be made and the morphology studied.
We can concluded that the emission is
consistent with Gaussians between 0.3--0.6 $^{\circ}$, which correspond to
13--26 kpc at a distance of 2.5~Mpc.  Changing the input
gamma-ray's spatial distribution from Gaussian to a rectangular
shape gave similar results, i.e., Fig.~\ref{fig:yield}
can be reproduced by a 0.4$^{\circ}$ rectangular spatial emission.  This
can be understood by the efficiency reduction due to the limited FOV. The
acceptance for our telescope was reduced from an offset angle of
0.5$^{\circ}$.
The electrons of GeV energy associated with NGC 253 was reported
by radio observations of
Hummel, Smith, and van der Hulst (\cite{hummel1984}) and
Carilli et al. (\cite{Carilli}). 
The size of the emission region is similar to our result.
An interpretation of this results can be found in Itoh et al.
(\cite{Itoh2003}). 

\section{Conclusions}

In this paper we have concentrated on technical details concerning our
observations and analysis.  Statistically significant signals of
gamma-rays from the nearby starburst galaxy NGC~253 have been detected.  
The differential flux shows a turnover below 0.5~TeV.  The spatial
distribution of the gamma-ray emission is broader than that of a
point-source.  This is consistent with a width of 0.3-0.6$^{\circ}$,
corresponding to 13-26~kpc at the location of NGC~253.  A 
more detailed physical interpretation is 
presented elsewhere (Itoh et al., \cite{Itoh2003}).

%

\begin{acknowledgements}
 
We thank Prof.\ T.G.\ Tsuru for help to analyze the multi-wavelength
observational results. This work was supported by a Grant-in-Aid for
Scientific Research by the Japan Ministry for Education, Science,
Sports, and Culture, Australian Research Council, and Sasagawa 
Scientific Research Grant from the Japan Science Society.
The support of JSPS Research Fellowships for A.A., J.K., K.O.\ and K.T.\
are gratefully acknowledged.
The Digitized Sky Survey was produced at the Space Telescope Science
Institute under U.S. Government grant NAG W-2166. The images of these
surveys are based on photographic data obtained using the Oschin
Schmidt Telescope on Palomar Mountain and the UK Schmidt
Telescope. The plates were processed into the present compressed
digital form with the permission of these institutions.

\end{acknowledgements}

\end{document}